\documentclass[twocolumn,noshowpacs,preprintnumbers,superscriptaddress,amsmath,amssymb,floatfix, groupedaddress,showkeys]{revtex4}
\usepackage[caption=false]{subfig}
\usepackage{graphicx}
\usepackage{color}
\usepackage{setspace}
\usepackage[title]{appendix}
\usepackage{placeins}
\usepackage{tikz}
\usepackage{mathtools}
\usepackage{amsthm}
\usepackage{color}
\usepackage{xkeyval,xcolor}
\usepackage{array}

\newcommand{\er}{Erd\H{o}s-R\'{e}nyi}

\DeclareFontFamily{OT1}{pzc}{}
\DeclareFontShape{OT1}{pzc}{m}{it}{<-> s * [1.10] pzcmi7t}{}
\DeclareMathAlphabet{\mathpzc}{OT1}{pzc}{m}{it}

\begin{document}
	
	\title{Epidemic spreading and control strategies in spatial modular network}
	\author{Bnaya Gross} 
	\email{bnaya.gross@gmail.com}
	\affiliation{Department of Physics, Bar-Ilan University, Ramat-Gan 52900, Israel}
	\author{Shlomo Havlin}
	\affiliation{Department of Physics, Bar-Ilan University, Ramat-Gan 52900, Israel}
	
	\begin{abstract}
		Epidemic spread on networks is one of the most studied dynamics in network science and has important implications in real epidemic scenarios. Nonetheless, the dynamics of real epidemics and how it is affected by the underline structure of the infection channels are still not fully understood. Here we apply the SIR model and study analytically and numerically the epidemic spread on a recently developed spatial modular model imitating the structure of cities in a country. The model assumes that inside a city the infection channels connect many different locations, while the infection channels between cities are less and usually directly connect only a few nearest neighbor cities in a two-dimensional plane. We find that the model experience two epidemic transitions. The first lower threshold represents a local epidemic spread within a city but not to the entire country and the second higher threshold represents a global epidemic in the entire country. Based on our analytical solution we proposed several control strategies and how to optimize them. We also show that while control strategies can successfully control the disease, early actions are essentials to prevent the disease global spread.
	\end{abstract}
	\maketitle

Network science is becoming one of the most fruitful research fields in the last decades explaining variety of phenomena in many complex systems such as the human brain \cite{moretti2013griffiths_critical_brain,sporns2010networks} the human microbiome \cite{smillie2011ecology,gibson2016origins,layeghifard2017disentangling}, protein-protein interactions \cite{kovacs2019network,de2015structural,li2017oncoppi}, climate \cite{yamasaki2008climate,jingfan2017network_climate,ludescher2014very_climate}, ecology \cite{paine1966food,polis1996food} and infrastructures \cite{yang2017small_power_grid,latora2005vulnerability_infrastructures,li-pnas2015}. Modelling of these systems and many others opened a new direction of studying many complex structures such as modular (community) networks \cite{palla2005uncovering,rosvall2008maps,gross2020interconnections,capocci2005detecting,shekhtman2015resilience_communities,newmanngirvan2002community}, multiplex networks \cite{nicosia2013growing,gomez2013diffusion,granell2013dynamical,bianconi2013statistical}, interdependent networks \cite{wang2013interdependent,buldyrev2010catastrophic,brummitt2012suppressing,baxter2012avalanche,gao2012networks,radicchi2013abrupt} and high order interactions networks \cite{lambiotte2019networks,de2020social,millan2020explosive}. These structures were studied under different processes and dynamics such as percolation \cite{bunde1991fractals, staufferaharony}, synchronization \cite{arenas2006synchronization,danziger2019dynamic,de2017diffusion}, reaction-diffusion \cite{weber2008pattern,cencetti2018pattern,lazaridis2018spontaneous,colizza2007reaction}, and epidemics \cite{pastor2015epidemic,bogua2003epidemic,wang2017unification}. \par
\begin{figure}
	\centering
	\begin{tikzpicture}[      
	every node/.style={anchor=north east,inner sep=0pt},
	x=1mm, y=1mm,
	]   
	\node (fig3) at (0,0)
	{\includegraphics[scale=0.7]{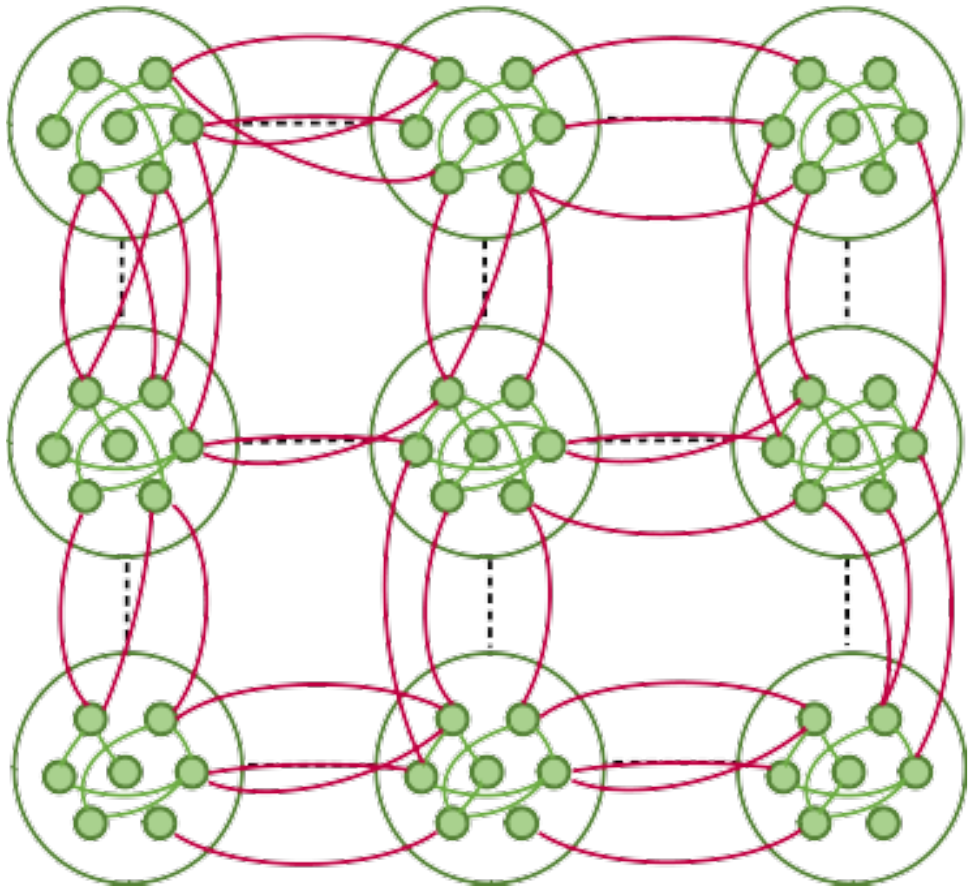}};
	\end{tikzpicture}
	\caption{\textbf{Illustration of the model.} The spatial modular model represents a structure of a network of infection channels inside cities (modules) and between cities. Inside a city, the infection channels are dense and spread randomly between different areas of the city (green links) like in an \er~network having random like structure while the infection channels from one city to another is usually possible between neighbouring cities (red links) having spatial like structure.}
	\label{fig:illustration}	
\end{figure}
When modeling a process, one should be careful not strictly fitting the model to a specific scenario which will reduce its generality, and to account for as many as possible of the important features of the process in order to make the model valid and useful in different scenarios. While recently the study of epidemic spread has been conducted on a community structure due to the human social organization \cite{palla2007quantifying,jin2001structure}, it mainly considered a random organization of the communities while neglecting the spatial structure  \cite{salathe2010dynamics,valdez2020epidemic,nadini2018epidemic,liu2005epidemic}. \par
In this paper, we applied the SIR model to study the epidemic spreading in a 2D spatial community network model \cite{vaknin2019spreading,gross2020two}, see Fig. \ref{fig:illustration}, to better describe epidemic spreading in human social community organization. Each community can represent a city and the entire network represents a country. We find that the epidemic spreading in such networks experience two epidemic transitions one at $\beta^{ER}_c$ when a local outbreak spread in the origin city but not in the entire country and the second at $\beta^{2D}_c$ when the epidemic spreads in the entire country. We find analytically the values of both epidemic thresholds and develop several control strategies and optimization methods to mitigate the spreading of the disease. Moreover, we show the importance of early actions and how delaying might result in a global spread of the epidemic with catastrophic results.

\underline{\textit{Model.--}} The spatial community model \cite{vaknin2019spreading,gross2020two} illustrated in Fig. \ref{fig:illustration}, represents the infection channels within and between communities on a 2-dimensional square lattice with $N=L \times L$ lattice sites, where $L$ is the linear size of the lattice and the lattice sites are the nodes of the network. The lattice is divided into smaller squares of linear size $\zeta$ representing communities, e.g., cities. The number of nodes in each community is $N_c= \zeta \times \zeta$. Thus, the number of communities in our model is $n = N/N_c=L^2 / \zeta^2$. We assume that inside a city the infection channels are dense and spread randomly between different sites in the city. Therefore, each community will be connected randomly like an \er~network (ER) with an average degree $k_{intra}$. In contrast, the infection channels between cities are less dense than within cities and usually connecting neighbouring cities. Thus, we assume that in addition to the intra-links linking the nodes in the same community, there are fewer inter-links which connect the nodes located in neighbouring communities. We assume that each node has inter-links distributed according to a Poisson distribution with the average degree $k_{inter} \ll k_{intra}$. Each inter-link is connected randomly to one of the nodes of the four nearest neighbouring communities occupying adjacent squares on the lattice as shown in Fig. \ref{fig:illustration}. This assumption represents the fact that roads or railways usually connect neighbouring cities. For brevity of notations, we denote $K\equiv k_{intra}$ and $Q\equiv k_{inter}\zeta^2$, where $Q$ is the average number of inter-links emanating from each community to its four neighbours. \par
\begin{figure}
	\centering
	\begin{tikzpicture}[      
	every node/.style={anchor=north east,inner sep=0pt},
	x=1mm, y=1mm,
	]   
	\node (fig3) at (0,0)
	{\includegraphics[scale=0.4]{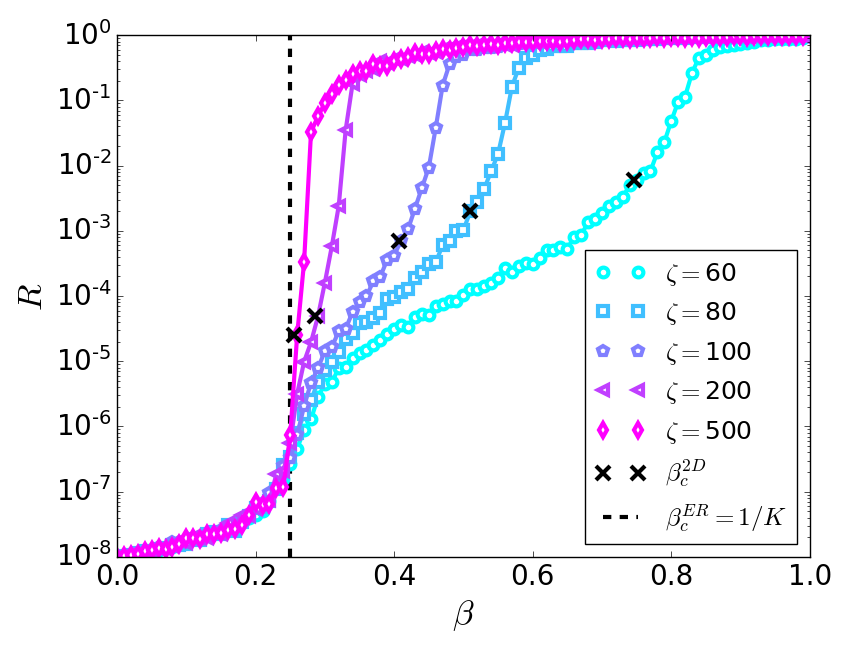}};
	\end{tikzpicture}
	\caption{\textbf{Two epidemic transitions.} Simulations of the epidemic recovered cluster $R$ as a function of $\beta$ for different values of $\zeta$ on a log-linear graph with $K = 4$ and $k_{inter} = 10^{-3}$. The epidemic recovered cluster is measured once no infected nodes remain. Two distinct epidemic transitions are observed. The first (lower) transition at $\beta^{ER}_c = 1/ K$ occurs when a small outbreak spread in a city but not in the entire country. The second (higher) transition at $\beta^{2D}_c$ when a global epidemic spread in the whole country is obtained from Eq. \eqref{eq:betac} and is denoted by black $\times$.  Here $N = L \times L \sim 10^8$ ($L = 9960$ for $\zeta = 60$ and $L = 10^4$ for the other $\zeta$ values).}
	\label{fig:generalI}	
\end{figure}
This model has two important limits. For $\zeta \to L$ the models generate an ER network while for $L \gg \zeta \to 0$ strong spatial (regular lattice) behaviour is observed. Moreover, for intermediate values of $L > \zeta > 0$ mean-field behaviour is observed in small scales (below $\zeta$) and spatial behaviour on large scales (above $\zeta$). Note that a similar but homogeneous model has been studied with similar limits \cite{danziger-epl2016,grossvaknin-JPS,vaknin-njp2017,bonamassagross2019critical}. However, due to its homogeneous structure (and not heterogeneous as in the present modular model) it experiences very different features compared to our model with a single epidemic transition.  \par
\underline{\textit{Analytical and numerical results.--}} We study the epidemic spread in our model using the SIR model.  We start with a single infected node in a random community while all other nodes are susceptible. At each time step every infected node attempts to infect its susceptible neighbours independently with infection probability $\beta$ and become recovered afterwards. The impact of the epidemic outbreak for different values of $\beta$ can be measured as the fraction of the recovered cluster, $R$, once there are no more infected nodes as shown in Fig. \ref{fig:generalI}. As expected, for $\zeta \to L$ the behaviour of the network approaches the behaviour of a regular ER with $\beta^{ER}_c = 1 / K$. It can be seen that for any value of $\zeta \gg 1$ the recovered cluster has two inflection points. The first (lower) transition at $\beta^{ER}_c$ when a local outbreak spreads within the origin city but does not propagate in the entire country. The second (higher) transition at $\beta^{2D}_c$ when the epidemic spreads in the entire country. These two epidemic transitions are analogues to the two percolation transitions found in Ref \cite{gross2020two} in the same model although the analytical solution is different. While the position of the first transition does not depend on $\zeta$, the position of the second decreases with $\zeta$, and at large $\zeta$ it almost coalesces with the first one. As we will see, this second transition corresponds to the bond percolation threshold of the spatial network of communities which has a topology of a square lattice. This is due to the known mapping between the SIR model and bond percolation \cite{grassberger1983critical,sander2002percolation}. Near this transition the epidemic spread globally in the country composed of infected cities and the size of their local outbreaks can be found analytically. Finally, near the first (lower $\beta_c$) transition corresponding to the epidemic threshold of ER network, the local outbreak disappears as well and the average recovered cluster swiftly goes to zero as $\beta$ decreases below $\beta^{ER}_c$. \par
\begin{figure*}
	\centering
	\includegraphics[scale=0.9]{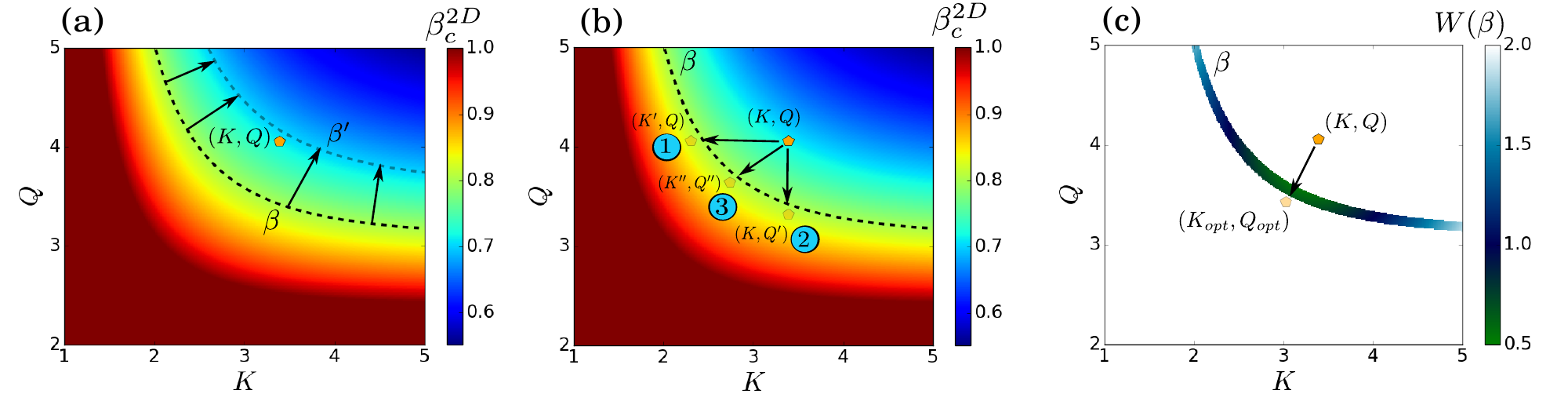}
	\caption{\textbf{Control strategies and optimization.} A given country (orange pentagon) is placed in the structural parameter space $(K,Q)$ with epidemic threshold $\beta^{2D}_c$ obtained from Eq. \eqref{eq:betac}. For an epidemic spread with infection probability $\beta$ it is desired to position the country in such a way that $\beta < \beta^{2D}_c(K,Q)$, such that there will be no epidemic. This can be achieved by the following strategies:   \textbf{(a) Social strategy.} Assume that the natural epidemic infection rate is $\beta = 0.8 > \beta^{2D}_c$ (thick black dashed line) above the epidemic threshold of the country. By using social distancing or mask-wearing the infection probability could be reduced to $\beta^\prime = 0.7 < \beta^{2D}_c$ (grey dashed line) and thus becomes below the epidemic threshold and the disease will not spread. \textbf{(b) Quarantine strategies.} By reducing the infection channels in and between the cities (reducing $K$ and $Q$ respectively) the country's position in the structural parameter space can be changed and the epidemic threshold will increase such that the infection probability will be below the epidemic threshold. Three ways are suggested: (1) local quarantine strategy within cities by reducing $K \to K^{\prime}$. (2) global quarantine strategy between cities by reducing $Q \to Q^\prime$. (3) Mixed strategy by reducing both $K \to K^{\prime\prime}$,$Q \to Q^{\prime\prime}$. \textbf{(c) Strategies optimization.} A weight function, $W(\beta)$, can be evaluated for optional locations for the parameters space of the country based on economical, health, and social arguments such that $\beta_c(K,Q) \to \beta^{+}$. Optimization of the weight function will yield the optimal location for the country $(K_{opt}, Q_{opt})$. Here we used the Euclidean distance in the parameters space as a weight function $W(\beta) = \sqrt{(K-K^{\prime \prime})^2 + (Q-Q^{\prime \prime})^2}$ and its optimization (minimizing) will yield the shortest Euclidean distance which represents minimal reduction of the inter and intra links.}
	\label{fig:phase}	
\end{figure*} 
Next, we demonstrate that the second inflection point (at higher $\beta$) corresponds to the bond percolation transition on a square lattice due to the mapping from SIR \cite{grassberger1983critical,sander2002percolation}. To this end we compute the position of the inflection points for different $\zeta$ analytically using the well known fact that the bond-percolation threshold for a square lattice is $1/2$ (See \cite{bunde1991fractals} and appendix B). Here we will use the bond percolation threshold value to find the value of $\beta^{2D}_c$ at which the epidemic spread in the entire country.  The probability that one of $Q$ inter-links emanating from a given community connects to one of its 4 neighbours is $1/4$. Therefore, the number $k$ of the inter-links connecting these two neighbouring communities is distributed with a binomial distribution $P_k(Q)=(1/4)^k(3/4)^{Q-k}C_Q^k$. The probability that a randomly chosen node will be part of the local outbreak in a community (city) is given by the epidemic component of ER network (See \cite{newman2002spread} and appendix A),
\begin{equation}
S = 1 - e^{-K \beta S} .
\label{eq:S}
\end{equation}
The spread of a local outbreak in a city to one of its neighbouring cities happens through the city's inter-links. Above $\beta^{ER}_c$ the local epidemic spreads in the whole city and the finite non-infected clusters are of size $s\ll \zeta^2$ and will have a very low chance to have more than one interlink for $s \cdot k_{inter}\ll 1$. 
Thus, assuming a very small $k_{inter}$, the probability that a local outbreak in a city will spread to one of its neighbours through a single inter-link is $S\beta$ and the probability that a local outbreak will not spread through one of the city's inter-links is
\begin{equation}
\beta_b=\sum_k P_k(Q)(1-S \beta)^k=\left [ \frac{3}{4} + \frac{1}{4}(1 - S\beta) \right ]^Q .
\label{eq:betab}
\end{equation}
At the lattice epidemic threshold, the probability that a local outbreak will spread to neighbouring cities, $\beta_b$ should be $1/2$, the bond percolation threshold. Thus, the lattice epidemic threshold, $\beta^{2D}_c$, where the epidemic spread in the entire country can be obtained using Eqs. \eqref{eq:S} and \eqref{eq:betab},
\begin{equation}
\beta^{2D}_c = \frac{4(1-2^{-1/Q})}{1 - \exp(-4K(1 - 2^{-1/Q}))} .
\label{eq:betac}
\end{equation}
In this case, the size of the local outbreak in the infected cities, $S(\beta^{2D}_c)$, is not zero as it is usually in second order phase transitions since $\beta^{ER}_c < \beta^{2D}_c$ and each infected city is above criticality. The size of the local outbreak at the lattice threshold can be found analytically directly from Eqs. \eqref{eq:S} and \eqref{eq:betab},
\begin{equation}
S(\beta^{2D}_c) = 1 - \exp(-4K(1 - 2^{-1/Q})) .
\label{eq:betac_giant}
\end{equation}
In the limit of $\zeta \to L$, Eq. \eqref{eq:betac_giant} takes the form
\begin{equation}
S(\beta^{2D}_c)  \simeq \frac{4K}{Q} \ln2,
\label{eq_G_pc_approx}
\end{equation}
and $\beta^{2D}_c = \beta^{ER}_c = 1/K$ as expected. \par

\underline{\textit{Control strategies and optimization.--}} When an epidemic spreads in a country it is essential to develop control strategies to tame the disease. For this purpose it is very insightful to study the phase diagram in the structural parameters space $(K,Q)$ (Fig. \ref{fig:phase}). A given country is placed in the structural parameter space according to its inter-degree $K$ and city intra-degree $Q$ for which its epidemic threshold can be calculated from Eq. \eqref{eq:betac}. For the case of an epidemic with infection probability $\beta > \beta^{2D}_c(K,Q)$ the epidemic will spread in the country and an appropriate control strategy should be considered. This scenario can be visually observed in the parameter space when the system is placed above the line $\beta = \beta^{2D}_c(K,Q)$ (Fig. \ref{fig:phase}a black dashed line). Since $\zeta$ is related to the spatial structure of the cities and rarely changes during the timescale of the epidemic, one should try effecting the other parameters in order to control the disease. The main goal is to achieve a state such that $\beta < \beta^{2D}_c(K,Q)$ and the epidemic will not spread. In such a case the system will be placed below the line $\beta = \beta^{2D}_c(K,Q)$ (Fig. \ref{fig:phase}a grey dashed line). This can be achieved through the following strategies: \par \noindent
\textbf{Social strategy.} Since the epidemic propagates through human interactions, the basic approach could be based on reducing the infection probability $\beta \to \beta^\prime$ in such a way that $\beta^{\prime} < \beta^{2D}_c(K,Q)$ as shown in Fig. \ref{fig:phase}a. This can be achieved by social distancing, mask-wearing, etc to reduce the probability of an infected person to infect others. This approach and its effectiveness vary between countries and populations due to many factors such as population dynamics. \par \noindent
\textbf{Quarantine strategy.-} In the case that social strategies are not effective enough and after applied still $\beta^{\prime} > \beta^{2D}_c(K,Q)$, one can apply quarantine strategies by reducing the infection channels in and between cities. The first approach reduces the degree within cities $K \to K^\prime$ such that $\beta < \beta^{2D}_c(K^\prime,Q)$ as shown in the first option in Fig. \ref{fig:phase}b. $K^\prime$ should be reduced below the critical value $K_c$ obtained from $\beta   =  \beta^{2D}_c(K_c,Q)$ which can be analytically found from Eq. \eqref{eq:betac},
\begin{equation}
K_c = -\frac{1}{4(1-2^{-1/Q})} \log \left [ 1 - \frac{4(1-2^{-1/Q})}{\beta}\right].
\label{eq:Kc}
\end{equation}
The second approach reduces the degree between cities $Q \to Q^\prime$ (through $k_{inter}$ since $\zeta$ is usually fixed) such that $\beta < \beta^{2D}_c(K,Q^\prime)$ as shown in the second option in Fig. \ref{fig:phase}b. The value of $Q^\prime$ should be reduced below the critical value $Q_c$ which can be graphically evaluated from Eq. \eqref{eq:betac}.\\
The third approach involve combining the above two options by reducing both the degree in and between cities $(K,Q) \to (K^{\prime\prime},Q^{\prime\prime})$ such that $\beta < \beta^{2D}_c(K^{\prime\prime},Q^{\prime\prime})$ as shown in the third option in Fig. \ref{fig:phase}b.\par
When considering which approach to adopt, an optimization method can be developed. To this end, a weight function $W(\beta)$ can be evaluated for optional locations for the country in the parameters space based on economical, health, and social arguments. $W(\beta)$ is evaluated on the  $\beta_c(K,Q) \to \beta^{+}$ line as shown in Fig. \ref{fig:phase}c. Optimization of the weight function will yield the optimal location for the country $(K_{opt}, Q_{opt})$ in the parameter space. In Fig. \ref{fig:phase}c we used the Euclidean distance in the parameters space as a weight function $W(\beta) = \sqrt{(K-K^{\prime \prime})^2 + (Q-Q^{\prime \prime})^2}$ and its optimization (minimizing) will yield the shortest Euclidean distance which represents the minimal reduction of the inter and intra links, i.e., minimal restrictions. However, in a real scenario much more complex function is required. This function should take into account the economic cost of reducing the degree in and between the cities, social cost of quarantine, and many other collateral damage factors.
\par
\begin{figure}
	\centering
	\begin{tikzpicture}[      
	every node/.style={anchor=north east,inner sep=0pt},
	x=1mm, y=1mm,
	]   
	\node (fig3) at (0,0)
	{\includegraphics[scale=0.53]{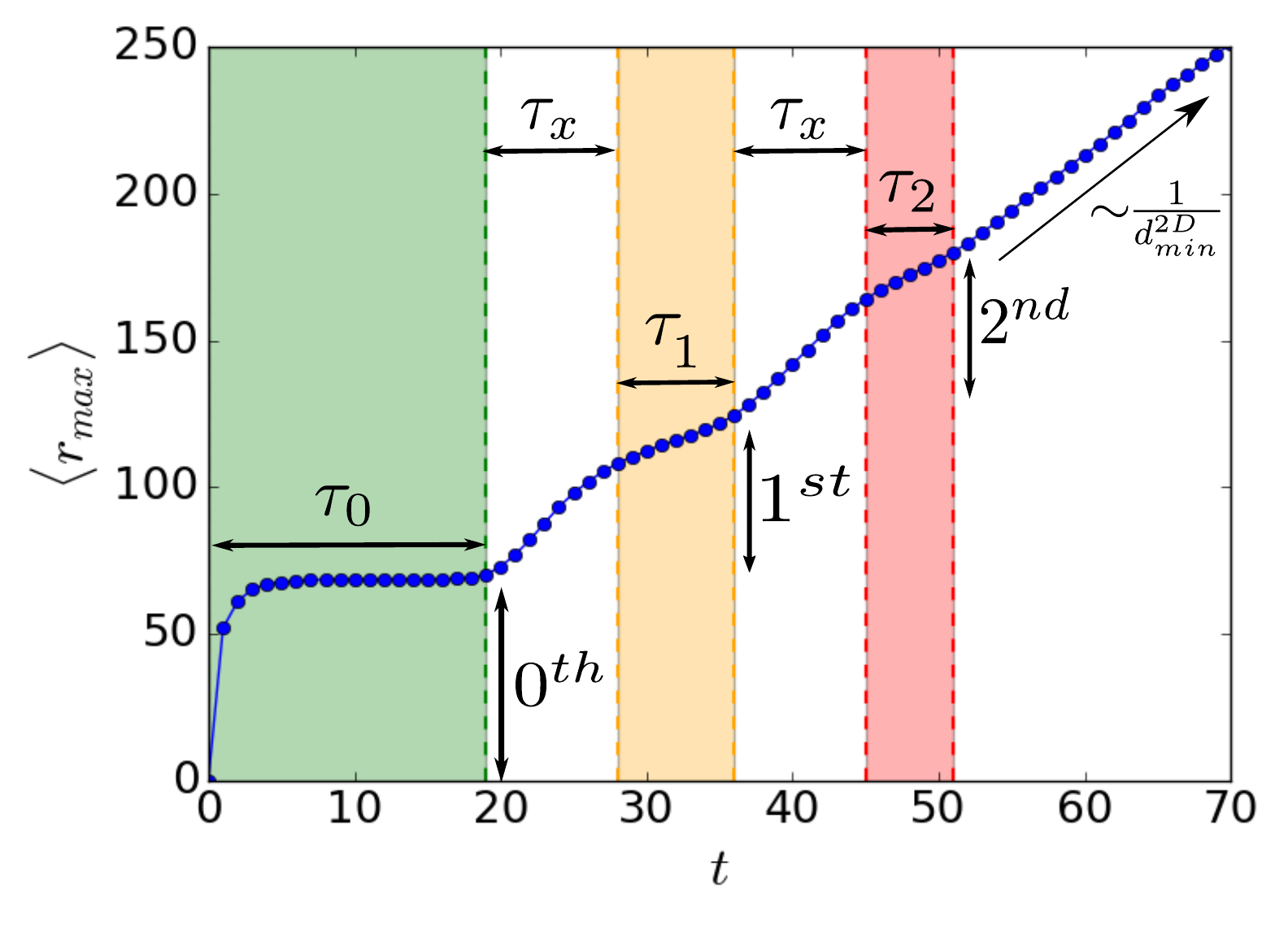}};
	\end{tikzpicture}
	\caption{\textbf{Epidemic spatial propagation.} The average maximum extent of the epidemic, $\langle r_{max} \rangle$, is measured as a function of time at $\beta^{2D}_c$. At early times the epidemic spread locally within the origin city (zeroth circle) for a period of time $\tau_0$ with a constant $\langle r_{max} \rangle \sim \zeta$. Afterwards, the epidemic spread to the first circle of cities around the origin city for a period of time $\tau_1$ and later to the second circle of cities for a period of time $\tau_2$. The transition time between the circles is denoted by $\tau_x$. As the epidemic evolves the distinction between circles decreases and identifying the distance of the disease from the origin is less clear. At later times the distinction of circles disappear completely and a clear spatial propagation is observed with $\langle r_{max} \rangle \sim t^{1/d^{2D}_{min}} = t^{1/1.13}$ \cite{bunde1991fractals}. The reason for the disappearance of the distinction between circles at later times is because the epidemic may spread faster in a given area and slower in another leading to inconclusive distinction between circles. Here we used $\zeta = 100$, $K = 4$, $k_{inter} = 10^{-3}$ and $\beta^{2D}_c = 0.407$.}
	\label{fig:r_t}	
\end{figure}
\begin{figure*}
	\centering
	\begin{tikzpicture}[      
	every node/.style={anchor=north east,inner sep=0pt},
	x=1mm, y=1mm,
	]   
	\node (fig3) at (0,0)
	{\includegraphics[scale=0.8]{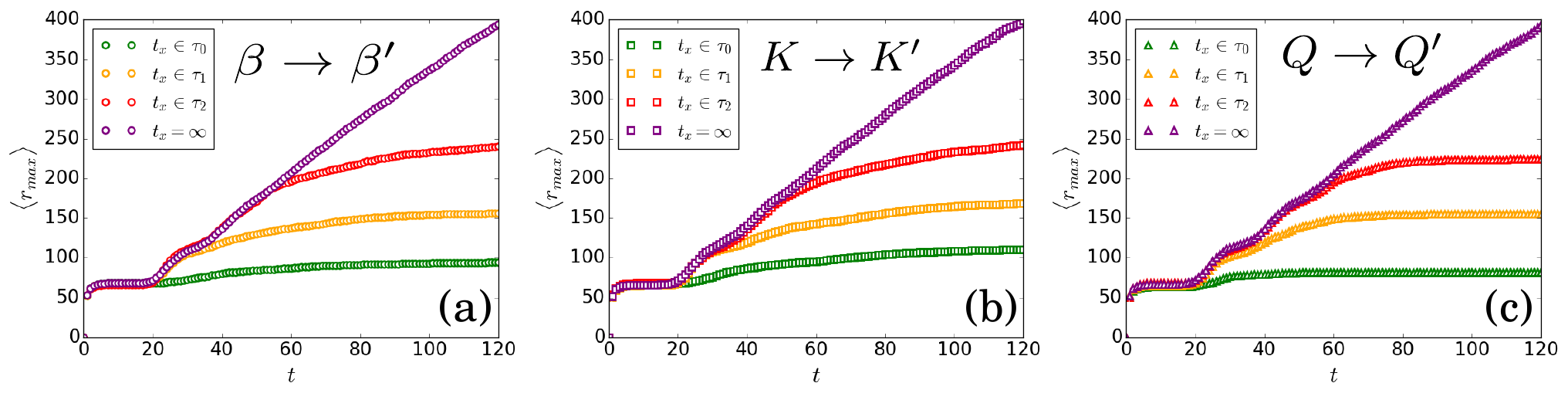}};
	\end{tikzpicture}
	\caption{\textbf{Intervention timing.} Here we show the effect of different control strategies performed at intervention timing $t_x$ at different circles' timescales corresponding to Fig. \ref{fig:r_t} and compare them to the scenario of no intervention corresponding to the case of $t_x = \infty$. The epidemic start spreading with the parameters $\zeta = 100$, $K = 4$, $k_{inter} = 10^{-3}$, $Q = 10$ and $\beta = \beta_c = 0.407$. \textbf{(a) Social strategy.} $\beta \to \beta^\prime = 0.3$. \textbf{(b) Quarantine strategy within cities.} $K \to K^\prime = 3$. \textbf{(c) Quarantine strategy between cities.} $Q \to Q^\prime = 1$ by reducing $k_{inter} \to k_{inter}^\prime = 10^{-4}$. The epidemic extent at $t_x$ is $\langle r_{max} \rangle_x$ and when the epidemic stops to spread it is $\langle r_{max} \rangle_f$. In all cases the intervention successfully stop the disease spatial propagation with $\langle r_{max} \rangle_f \approx \langle r_{max} \rangle_x$.}
	\label{fig:control}	
\end{figure*}
\underline{\textit{Consequences of late intervention.--}} While well-performed control strategies (Fig. \ref{fig:phase}) will result in epidemic extinction, the timing of the intervention plays a significant role. In many cases, early action can control the disease rapidly with a low amount of infections while late reply may not be efficient since the epidemic may already spread globally. To understand the effect of intervention timing we study the spatial propagation of the epidemic as a function of time. In Fig. \ref{fig:r_t} we show the average maximal extent of the epidemic from the origin, $\langle r_{max} \rangle$, as a function of time at $\beta^{2D}_c$. Assuming a small $k_{inter}$, at early times the spatial structure of cities around the origin city can be observed in the spatial propagation of the diseases. In the beginning, the disease spread locally within the origin city (zeroth circle) with $\langle r_{max} \rangle \sim \zeta$. The timescale of this stage is $\tau_0$ which generally depends on $\zeta$, $K$, and $k_{inter}$. Afterwards, the epidemic spread in the first circle of cities around the origin city with timescale $\tau_1$ and later in the second circle with timescale $\tau_2$. $\tau_x$ is the timescale of crossing between circles. Interestingly the timescale of the diseases spread in each circle gets shorter as the the disease spread further and eventually completely disappear and a clear 2D spatial propagation (of fractal type-since the system is at criticality) is observed with $\langle r_{max} \rangle \sim t^{1/d^{2D}_{min}} = t^{1/1.13}$ \cite{bunde1991fractals} as shown in Fig. \ref{fig:r_t}. The reason for the disappearance of the distinction between circles at later times is because the epidemic may spread faster in a given area and slower in another leading to inconclusive distinction between circles.
\par
This phenomenon of spatial identification of the disease in different circles at early times and its disappearance at later times has significant consequences for intervention timing. At early times when the epidemic can be identified in a given circle, quarantine strategies (Fig. \ref{fig:phase}b) can be applied locally around the circle and control the disease without affecting the whole country (as could have been done around Hubei province in China in the case of COVID-19 \cite{gross2020spatio}). In contrast, in later times the identification of the disease location becomes inconclusive and intervention should be applied on much larger scales to control the disease. Moreover, while in the close circles the spatial identification is valid as shown in Fig. \ref{fig:r_t} for the first and the second circles, the time window for action gets shorter and shorter ($\tau_0 > \tau_1 > \tau_2$) and require fast decisions. In Fig. \ref{fig:control} we tested different control strategies discussed in Fig. \ref{fig:phase} applied in different timing, $t_x$. In order to test if the intervention was successful, it is of interest to compare $\langle r_{max} \rangle_x$, the epidemic extent at $t_x$, and the epidemic extent once the epidemic stop spreading $\langle r_{max} \rangle_f$. As shown in Fig. \ref{fig:control} the intervention was successful in controlling the disease with  $\langle r_{max} \rangle_f \approx \langle r_{max} \rangle_x$. Nonetheless, while the control strategies at early times successfully stop the disease propagation, the extent of action required to control the disease increases as the intervention is performed later resulting with a larger amount of infections as the epidemic spreads further. These results highlight the importance and impact of early actions.

\par
\underline{\textit{Acknowledgements.--}} We thank the Israel Science Foundation, ONR, the BIU Center for Research in Applied Cryptography and Cyber Security, NSF-BSF Grant no.\ 2019740, and DTRA Grant no.\ HDTRA-1-19-1-0016 for financial support.
\begin{appendices}
	\numberwithin{equation}{section}
	\numberwithin{figure}{section}	
\section{Epidemic spread in ER networks}
\begin{figure}
	\centering
	\begin{tikzpicture}[      
	every node/.style={anchor=north east,inner sep=0pt},
	x=1mm, y=1mm,
	]   
	\node (fig3) at (0,0)
	{\includegraphics[scale=0.4]{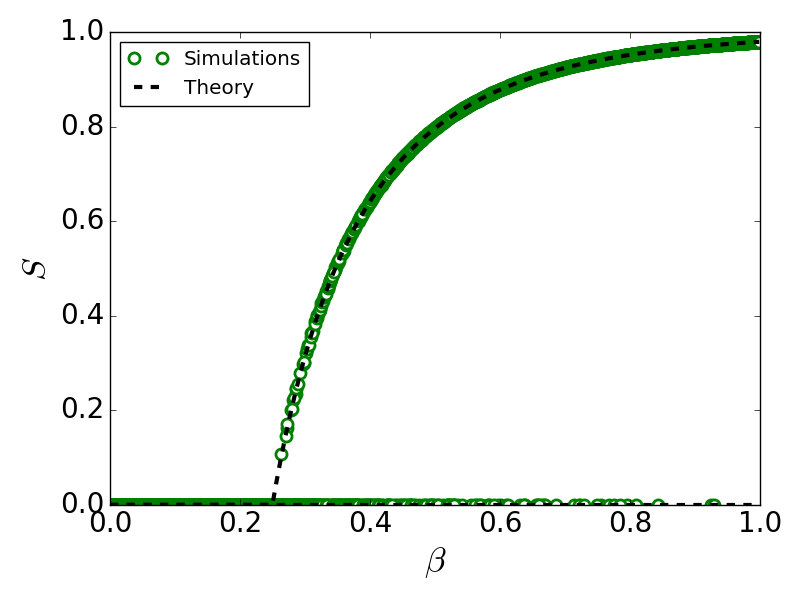}};
	\end{tikzpicture}
	\caption{\textbf{Epidemic spread in ER networks.} Theory is obtained from Eq. \eqref{eq:S_app} and simulation performed with  $N = 10^6$ and $K = 4$. The epidemic threshold is $\beta^{ER}_c = 1/K = 1/4$.}
	\label{fig:ER}	
\end{figure}

To develop an analytical solution for epidemic spread in ER networks we will follow the formalizm developed by Newman \cite{newman2002spread}.
We will use the generating functions:
\begin{align}
G_0(u) = \sum_k p_k u^k\;,\label{eq:G0}\\
G_1(u) = \frac{1}{K}G^\prime_0(u) .\label{eq:G1}
\end{align}
for the degree distribution and the outgoing edges distribution respectively. In order to find the the size of the epidemic outbreak we need the generating functions for the distribution of the \textit{occupied edges}. Thus, following Ref. \cite{newman2002spread} the generating function for the occupied edges for epidemic with infection probability $\beta$ is:
\begin{equation}
G_0(u;\beta) = G_0(1 + (u-1)\beta).
\end{equation}
and
\begin{equation}
G_1(u;\beta) = G_1(1 + (u-1)\beta).
\end{equation}
For the case of ER networks $p_k = \frac{K^ke^{-K}}{k!}$, thus:
\begin{equation}
G_0(u;\beta) = G_1(u;\beta) = e^{-K\beta(1-u)} 
\end{equation}
where $K$ is the average degree. The size of the epidemic outbreak is $S(\beta) = 1- G_0(u;\beta)$ and $u(\beta) = G_1(u;\beta)$. Thus, $S = 1-u$ and a self consistent equation can be written:
\begin{equation}
S = 1 - e^{-K \beta S}
\label{eq:S_app}
\end{equation}
with the epidemic threshold $\beta^{ER}_c = 1/K$. Theory and simulation show excellent agreement as shown in Fig. \ref{fig:ER}.
\par
\section{Epidemic spread in 2D square lattice}
\begin{figure}
	\centering
	\begin{tikzpicture}[      
	every node/.style={anchor=north east,inner sep=0pt},
	x=1mm, y=1mm,
	]   
	\node (fig3) at (0,0)
	{\includegraphics[scale=0.4]{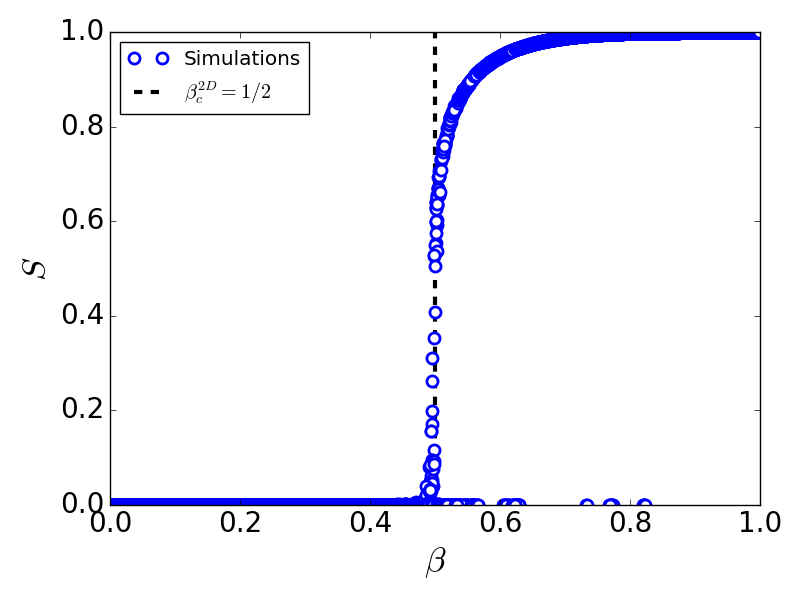}};
	\end{tikzpicture}
	\caption{\textbf{Epidemic spread in 2D square lattice.} The simulations were performed for $N = L \times L$ with $L = 1000$. The epidemic threshold is $\beta^{2D}_c = 1/2$.}
	\label{fig:SL}	
\end{figure}

The mapping between bond percolation and the SIR model yield the same epidemic threshold for 2D square lattice $\beta^{2D}_c = 1/2$ \cite{grassberger1983critical,sander2002percolation}.
Fig.~\ref{fig:SL} show simulations of the SIR model on a 2D square lattice with the epidemic threshold $\beta^{2D}_c = 1/2$ as expected.
\end{appendices}
\bibliographystyle{unsrt}

\end{document}